# EEG-to-Voice Decoding of Spoken and Imagined speech Using Non-Invasive EEG

Hanbeot Park[1], Yunjeong Cho[1], Hunhee Kim[1,2,*]

[1]Department of Data Engineering, Pukyong National University, South Korea

[2]Department of Computer Engineering and Artificial Intelligence, Pukyong National

University, South Korea

[*]Corresponding Author: h2kim@pknu.ac.kr

## Abstract

Restoring speech communication from neural signals is a central goal of brain–computer interface research, yet EEG-based speech reconstruction remains challenging due to limited spatial resolution, noise susceptibility, and the absence of temporally aligned acoustic targets in imagined speech. In this study, we propose an EEG-to-Voice paradigm that directly reconstructs speech from non-invasive EEG signals without dynamic time warping (DTW) or explicit temporal alignment. The proposed pipeline generates Mel-spectrograms from EEG in an open-loop manner using a subject-specific Generator, followed by pretrained vocoder and automatic speech recognition (ASR) modules to produce speech waveforms and decode text. Separate Generators were trained for spoken speech and imagined speech, and transfer learning–based domain adaptation was applied by pretraining on spoken speech and adapting to imagined speech. A minimal language model–based correction module was optionally applied to correct limited ASR errors while preserving semantic structure. The framework was evaluated under 2-s and 4-s speech conditions using acoustic-level metrics (PCC, RMSE, MCD) and linguistic-level metrics (CER, WER). Stable acoustic reconstruction and comparable linguistic accuracy were observed for both spoken speech and imagined speech. While acoustic similarity decreased for longer utterances, text-level decoding performance was largely preserved, and word-position analysis revealed a mild increase in decoding errors toward later parts of sentences. The language model–based correction consistently reduced CER and WER without introducing semantic distortion. These results demonstrate the feasibility of direct, open-loop EEG-to-Voice reconstruction for spoken speech and imagined speech without explicit temporal alignment.

# Introduction

Human language serves as a fundamental and powerful means of expressing thoughts and emotions, playing a central role in everyday life and social interactions. However, when neurological damage such as stroke or amyotrophic lateral sclerosis occurs, speech production abilities may be partially or completely lost, severely restricting an individual's communicative autonomy and quality of life[1]. In this context, brain–computer interface (BCI) research aimed at restoring communication through neural signals has been continuously developed over a long period of time. While early BCI studies primarily focused on simple command recognition or choice-based control, recent research has expanded toward directly reconstructing one of the highest-level human cognitive functions—linguistic thought—from neural signals[2–6]. Imagined speech, which does not involve overt articulation, has attracted attention as a task in which linguistic intent is internally formed without muscle movement or vocal output, thereby preserving the possibility of communication even under conditions where physical speech is impaired. These characteristics suggest the potential to recover linguistic information using only non-invasive neural signals and highlight the prospective value of providing a practical means of communication for patients with limited speech abilities.

Nevertheless, attempts to directly reconstruct natural, speech-level audio from non-invasive EEG signals still face several fundamental limitations. EEG signals suffer from limited spatial resolution and attenuation as they pass through the skull and surrounding tissues, and they are also highly susceptible to various sources of physiological noise, such as electromyographic activity and eye movements[7]. In addition, spoken speech and imagined speech exhibit inherent differences in terms of the presence of overt articulation, muscle activity, and the temporal clarity of speech onset, making it difficult to directly apply the same approaches to both conditions. Under imagined speech conditions, the absence of an actual acoustic signal inevitably gives rise to the problems of missing ground truth and temporal alignment uncertainty. For these reasons, existing EEG-based decoding studies have often been confined to limited word-level classification or phoneme-level decoding, and approaches relying on dynamic time warping (DTW) have been widely adopted to alleviate temporal alignment issues[5,8].

Building upon the limitations of existing approaches, this paper proposes an EEG-to-Voice paradigm that directly reconstructs speech from EEG signals without relying on a DTW-based alignment process. The proposed pipeline is designed as an integrated architecture that combines a generator for producing Mel-spectrograms from EEG signals with pretrained modules that transform the generated Mel-spectrograms into speech waveforms and text. Through this framework, speech corresponding to the speech duration is reconstructed in an open-loop manner, and the stability of the reconstructed speech is systematically evaluated in terms of both acoustic and linguistic information.

In addition, considering the differences in signal characteristics between spoken speech and imagined speech, separate generators were trained for each task. To mitigate the constraints of the imagined speech condition, in which no actual acoustic signal is available, a transfer learning–based domain adaptation strategy was employed, in which weights pretrained on spoken speech data were transferred to the imagined speech setting. Furthermore, to compensate for residual errors that may arise during the conversion of reconstructed speech into text via automatic speech recognition (ASR), a language model–based correction module was optionally applied, performing minimal linguistic adjustments without excessively altering the underlying semantic structure.

The objective of this study is to examine to what extent acoustic and linguistic information can be stably reconstructed for both spoken speech and imagined speech in a non-invasive EEG setting, despite differences in speech production and speech length across conditions. To this end, this work goes beyond assessing mere acoustic reconstructability and systematically analyzes whether trial-level speech reconstruction is preserved under conditions with varying speech lengths, how decoding accuracy evolves over time, and to what degree the reconstructed outputs retain meaningful information at the linguistic level. Furthermore, by taking into account the error characteristics that arise during ASR-based text decoding, we additionally evaluate the impact of language model–based correction on linguistic accuracy, thereby providing a comprehensive discussion on whether the proposed EEG-to-Voice pipeline can function as a practical means of linguistic information recovery across diverse conditions, including imagined speech.

# Method

Participants

A total of 23 participants (20 males and 3 females) were enrolled in this study. All participants were adults in their 20s to 30s, with no history of psychiatric disorders and no impairments in auditory, language, or motor functions. All participants were native speakers of Korean. The experiment was conducted in a shielded room, where each participant was seated alone in front of a monitor. Only individuals who did not report claustrophobia were included. As each experimental session required a high level of concentration for approximately 1–2 hours, individuals who were sleep-deprived or had consumed excessive alcohol prior to the experiment were excluded from participation. In addition, participants for whom stable contact between the scalp and EEG electrodes could not be ensured—such as those whose hair length made it difficult to wear the EEG equipment—were excluded.

Prior to participation, all participants were fully informed of the purpose and procedures of the experiment, as well as any potential discomfort, and provided written informed consent. Participants were informed that they could withdraw from the study at any time without penalty. The study protocol was approved by the Institutional Review Board (IRB) of Pukyong National University (approval number: PKNU 2024-12-001-002). Before the experiment, all participants underwent a hearing screening and a brief behavioral assessment to confirm that they met the eligibility criteria for participation.

EEG Recording

Neural signals were recorded using a wet-type 130-channel high-density EEG system manufactured by EGI (Electrical Geodesics Inc.). To reduce contact impedance between the electrodes and the scalp, the electrodes and scalp were thoroughly wet with a KCl solution prior to recording. The impedance of all electrodes was adjusted to be below 50 kΩ before the start of data acquisition. To prevent degradation of signal quality during recording, participants were instructed to minimize body and head movements as much as possible throughout the experiment.

Electrodes were arranged according to a geodesic montage uniformly distributed over the entire scalp, and data were sampled at 250 Hz. The Cz electrode was used as the reference. During recording, electrodes whose impedance remained above 50–100 kΩ were considered bad channels and were recorded separately in a text file.

To enable precise segment delineation, a GUI-based protocol server was integrated with the neural signal recording server. At the onset and offset of each task, the protocol server simultaneously transmitted annotations and CPU time via a multithreaded process, with annotations stored on the neural signal recording server and CPU time recorded on the protocol server. To ensure accurate temporal alignment, CPU time was additionally recorded at a sampling rate of 250 Hz. All subsequent signal processing and analysis procedures were performed offline after data acquisition.

Neural signal processing

First, bad channels identified during neural signal recording were removed as they contained noisy signals. Next, EEG recordings may include long-term upward or downward trends caused by factors such as electrode contact conditions, skin potential changes and slow physiological fluctuations which can lead to misinterpretation and degraded performance in EEG-based systems[9]. To address this issue, detrending was applied to each channel to remove baseline drift.

Because neural signal amplitudes and ranges vary substantially across subjects, directly inputting raw signals into the analysis algorithm may cause data from certain subjects to exert disproportionate influence. To mitigate this effect, electrode-wise min–max normalization was applied within each session to rescale the data to a common range, thereby improving the stability of model training.

Neural signals were band-pass filtered in the range of 0.1–100 Hz using a finite impulse response (FIR) filter implemented in the MNE-Python library. This frequency range was selected to remove low-frequency baseline drift below 0.1 Hz and high-frequency noise above 100 Hz while preserving the major EEG frequency bands required for analysis ranging from δ(0.5–4 Hz) to γ(30–100 Hz). Previous studies have reported that in imagined speech decoding, low-frequency power and cross-frequency dynamics provide critical information for phonemic and vowel representations and that high-frequency power contains important information not only for spoken speech but also for imagined speech[10]. In addition, prior work has demonstrated that localized activity in the low-gamma band during deep BCI learning is associated with improved BCI control performance[11]. Based on these findings and considering that high-frequency bands may convey meaningful information in high-level cognitive tasks a 0.1–100 Hz band-pass filter was employed in this study.

Because power-line interference is a well-known source of contamination in EEG/MEG analyses, notch filtering was applied at 60 Hz (and 120 Hz) to suppress power-line noise. After band-pass and notch filtering, physiological artifacts such as eye blinks (EOG) and electromyographic (EMG) activity were detected and removed using an independent component analysis (ICA)–based approach. ICA has been validated by numerous prior studies and widely used toolkits as an effective method for separating and removing noise components while preserving underlying neurophysiological signals[12].

### Signal Segmentation
To accurately extract task-relevant neural signal segments, annotations and CPU time information recorded alongside the neural and speech signals were matched to delineate task intervals. After preprocessing, neural signals and the corresponding raw speech signals were segmented using identical temporal intervals and the integrity of the speech signals within each segment was first verified. Subsequently, it was confirmed whether the speech signal within each segment correctly matched the given word or sentence label. When appropriate speech–label matching was achieved, annotations and CPU time information were further compared to verify that the neural signal segments were precisely cropped.
During preliminary analyses, we observed that inaccurate temporal alignment between EEG and speech signals led to unstable training outcomes even when using identical model architectures and training settings. Accordingly, a dual-validation procedure was employed in this study to ensure accurate temporal alignment. Through this process, segments in which neural and speech signals were precisely aligned to the task execution period were obtained.

### Voice signal processing
Speech signals were recorded simultaneously with neural signals during the spoken speech task at a sampling rate of 44.1 kHz. The recorded signals were subsequently resampled to 22.05 kHz and a noise reduction algorithm was applied to suppress environmental noise. The processed speech signals were then used for subsequent analyses. For time–frequency analysis, spectra were computed using a 1,024-point short-time Fourier transform (STFT) with a Hann window and a hop length of 256. Mel-spectrograms were generated by applying an 80-channel Mel filter bank over the 0–8 kHz frequency range. To ensure numerical stability, the amplitude spectra was transformed using $\log(\max(x, 10^{-5}))$ operation.

### Experimental paradigm

This study included two types of open-loop speech tasks to develop and evaluate the speech synthesis paradigm. All experimental sessions were conducted in a shielded room within the laboratory. Prior to the experiment, participants were verbally screened to confirm their suitability for the study followed by an explanation of the protocol and relevant instructions. When experiments were conducted across different days, particular care was taken during EEG setup to ensure consistent electrode placement such that the Cz channel was positioned at the midpoint between the mastoids. Each session consisted of a protocol in which a specific task was repeated for a total of 400 trials.

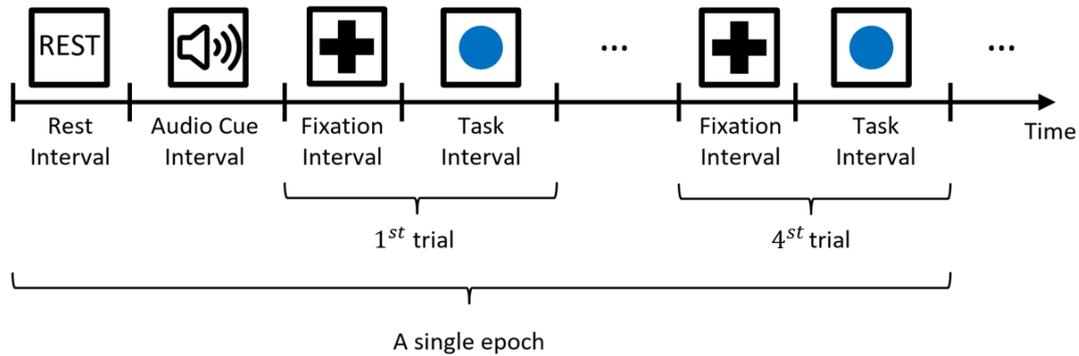

**Fig. 1| Experimental protocol.** Each experimental session begins with a rest interval, followed by the presentation of an auditory stimulus (audio cue) indicating the word or sentence task to be performed. After cue presentation, a 1-s fixation interval is provided to facilitate focused attention prior to task execution followed by the task interval during which the actual task is performed. Within a single block, the fixation interval and the task interval are each repeated four times. After completion of the block, the session transitions back to a rest interval after which a cue corresponding to a different word or sentence class is presented.

Fig. 1 illustrates the experimental protocol used in this study. Each block begins with a 3-s rest interval, during which the text cue "Rest" is displayed on the screen and participants are instructed to prepare for the next task while minimizing movement. This is followed by a cue interval lasting 2–4 s, during which an auditory cue corresponding to a word or sentence is presented together with a blank screen.

Participants are instructed to memorize the presented word or sentence for subsequent speech production. After the cue interval, a 1-s fixation interval is presented during which participants fixate on a black cross and prepare for speech. Immediately following the fixation interval, the task interval begins and participants are instructed to either overtly produce or imagine producing the word or sentence presented during the cue interval for a duration of 2–4 s. To maintain visual fixation during speech production a blue dot is displayed on the screen throughout the task interval.

To prevent the Generator from becoming biased toward excessively long silent segments a progress bar was displayed at the bottom of the screen to encourage participants to articulate slowly and clearly throughout the entire speech duration. The combination of the fixation interval and the task interval was repeated four times within a single epoch after which the next epoch began.

For each epoch, a cue corresponding to a different word or sentence class was presented and a total of 10 classes were cycled in a randomized order. Specifically, over 10 epochs, all word and sentence classes were each presented once and in the subsequent 10 epochs the same set of classes was repeated in a newly randomized order.

Based on this trial structure, participants performed the following two speech tasks:
(1) Spoken Speech: a task in which the presented word or sentence was overtly produced.
(2) Imagined Speech: a task in which the presented word or sentence was imagined without overt vocalization.

Prior to the main experiment, participants completed a practice session in which they performed both tasks after listening to the auditory cues. Through this process, the absence of impairments in auditory, language, and motor functions was confirmed and participants were equipped with the EEG apparatus. For both tasks, participants were instructed to articulate as clearly as possible to sustain speech throughout the task interval and to maintain a similar speech rate across the two tasks.

Accordingly, participants were instructed to initiate speech immediately at the onset of the task interval. Under the 2-s speech conditions, they were required to continue speaking until the progress bar was completed. Under the 4-s speech conditions, participants were instructed to continue speaking until the midpoint of the progress bar for word tasks and until the completion of the progress bar for sentence tasks.

For the imagined speech task, participants were explicitly instructed not to simply recall the word or sentence or imagine a scene but to imagine performing subvocal speech without overt muscle movements.

2-s speech condition

Under the 2-s speech condition, a total of 11 participants (9 males and 2 females) took part in the experiment and a total of 10 words and sentences were used as stimuli. The stimuli consisted of "go", "home", "time", "today", "wait", "go to school", "I am home", "Today is Sunday", "Wait a second", and "What time is it".

For each task (spoken speech and imagined speech), the experimental duration was approximately 1 hour and 30 minutes and the two tasks were performed consecutively on the same day. Each task consisted of a single session and each session comprised 400 trials.

4-s speech condition

Under the 4-s speech condition, a total of 12 participants (11 males and 1 female) participated in the experiment and a total of 20 words and sentences were used as stimuli. The stimuli were divided into two sets of 10 items and experiments were conducted on separate days for each set. On each day, both the spoken speech task and the imagined speech task were performed.

The duration of each task was approximately 2 hours. Each task consisted of two sessions, and a total of 800 trials were collected per task.

To maintain participants' alertness during the experiment, an experimenter monitored participants via a video camera from outside the soundproof booth. If a participant was judged to be drowsy, the booth was gently tapped during the rest interval to prompt alertness. If this situation occurred three times within the same session, the session was terminated and restarted.

Dataset

In the spoken speech task, speech signals and neural signals with clearly aligned speech onset and duration were collected simultaneously. In contrast, although the experimental procedure and stimulus conditions of the imagined speech task were nearly identical to those of the spoken speech task, no temporally aligned speech signals were available because no overt speech production was performed. During experiment design, participants were instructed to perform the imagined speech task by recalling the speech onset timing and speech rate used in the spoken speech task. Accordingly, for the imagined speech task speech recordings obtained from the spoken speech task for the same sentences were used as the target voice.

Both neural and speech data used in this study were collected directly by the authors, and no data augmentation was applied. The dataset was randomly split into training, validation, and test sets at a ratio of 2:1:1, and all experiments were conducted using subject-specific training. Each split was constructed to include all word and sentence classes to avoid the presence of unseen classes and the number of classes was evenly distributed across the training, validation, and test sets.

EEG-to-Voice pipeline

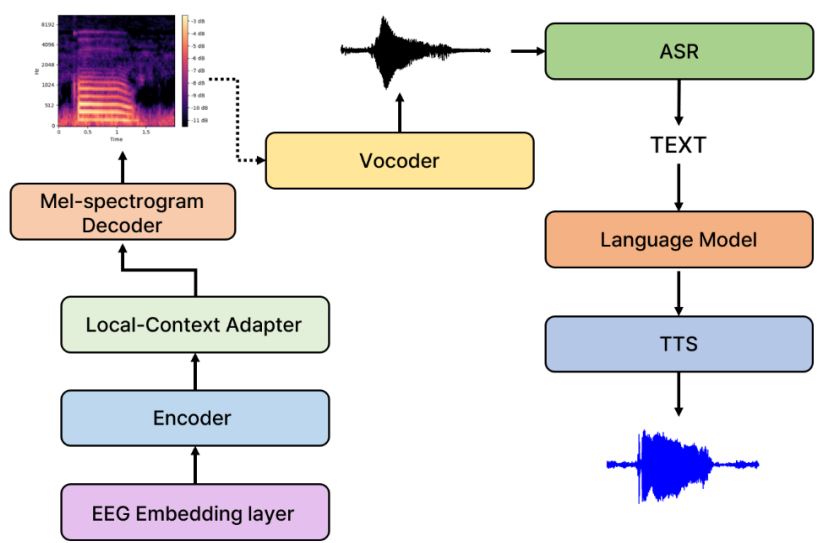

**Fig. 2| EEG-to-Voice pipeline for voice generation from EEG input.** Preprocessed EEG signals are directly used as input data without an explicit feature extraction stage. The EEG signals are passed through an EEG embedding layer and then fed into a Generator composed of an encoder a local-context adapter, and a Mel-spectrogram decoder, through which temporally aligned Mel-spectrograms are generated from the input EEG. The generated Mel-spectrograms are then provided to a pretrained vocoder to synthesize speech waveforms. Subsequently, the synthesized speech is converted into text via an automated speech recognition (ASR) module, and the text is refined through a correction module consisting of a language model and text-to-speech (TTS), resulting in the final personalized voice output. This pipeline enables continuous regression from neural signals to speech representations while preserving temporal correspondence.

In this study, an integrated pipeline was constructed to generate speech from EEG signals as input (Fig. 2). The pipeline consists of an EEG-based Mel-spectrogram generation model that sequentially produces Mel-spectrograms conditioned on the input EEG signals and pretrained modules that derive speech waveforms and text from the generated Mel-spectrograms. For the generated speech waveforms and text, a Correction Module can be optionally applied for post-processing.

The collected EEG signals, after preprocessing, are fed into a Generator composed of an EEG embedding layer, an encoder, a Local-context Adapter and a Mel-spectrogram decoder. The Generator does not employ attention-based explicit alignment mechanisms or trial-level global representations; instead, it learns temporal correspondence between the input EEG signals and the output Mel-spectrograms by leveraging local temporal context within the EEG time series. In addition, an open-loop architecture is adopted in which the generated outputs are not reused as inputs to the Generator, enabling the mapping between EEG signals and acoustic representations to be learned without speech feedback.

When preprocessed EEG signals are provided as input to a subject-specific Generator trained independently for each participant, Mel-spectrograms are generated and subsequently converted into 16 kHz speech waveforms through a pretrained vocoder. The generated speech is then decoded into text using a pretrained ASR model. All generation processes in this pipeline are performed offline without real-time constraints and both speech and text are produced based on the entire EEG input sequence.

The Correction Module consists of a pretrained language model and a subject-specific TTS system. After applying limited-scope correction to the decoded text the corrected text is resynthesized into personalized speech using the TTS module. The detailed configuration and application procedure of the Correction Module are described in the following section.

Pretrained Components

In this study, three pretrained models were employed. The vocoder and ASR modules were used to enable explicit speech generation from neural signals and subsequent conversion of the generated speech into text. First, the vocoder was used to synthesize high-resolution speech waveforms from Mel-spectrograms. Specifically, the Universal v1 model of HiFi-GAN was adopted. HiFi-GAN is a high-quality speech synthesis model based on adversarial training, and the Universal v1 model is a general-purpose version trained on many speakers and diverse speech datasets. Owing to its strong generalization performance on unseen datasets this model is well suited for the speech synthesis process in the present study[13].

Second, the ASR module was used to convert the generated speech into text. In this study, the pretrained HuBERT (Hidden-Unit BERT) model released by Facebook AI Research was employed[14]. HuBERT is a self-supervised learning–based speech representation model that demonstrates strong performance in speech-to-text mapping. Through this model, speech generated from neural signals could be effectively transformed into text sequences.

Third, a language model (LM) was used to correct the ASR outputs. Because ASR results may occasionally contain typographical errors or incomplete words an instruction-tuned language model was employed in this study to refine grammatical errors and improve semantic consistency. The LM adopted a prompt-based correction strategy and details regarding prompt design and system role configuration are provided in the Large Language Model settings.

Finally, the vocoder, ASR, and language model were all used in a frozen state without additional fine-tuning. These pretrained models served as foundational modules within the pipeline for speech synthesis, speech-to-text conversion and text correction.

Training Loss Term

Loss terms were defined to train the Generator, excluding the pretrained vocoder, ASR model, and language model. Two loss components were used for training the Generator: Reconstruction Loss $L_{rec}$ and CTC Loss $L_{ctc}$. The overall loss function is defined as follows:

$$L(G) = \lambda_{g1} L_{rec} + \lambda_{g2} L_{ctc}$$

Here, the weighting coefficients of the loss terms used for training the Generator were adjusted depending on the subject and the speech condition.

Reconstruction Loss

The reconstruction loss encourages the Generator to produce Mel-spectrograms that are like the target Mel-spectrogram by minimizing the discrepancy between the target and the generated outputs. In this study, the mean absolute error (MAE, L1 loss) was employed as the reconstruction loss:

$$L_{rec} = \frac{1}{N} \sum_{i=1}^{N} |\widehat{M_i} - M_i|$$

Let $M_i$ denote the $i$-th frame of the target Mel-spectrogram and $\widehat{M_i}$ denote the $i$-th frame of the Mel-spectrogram generated by the Generator, where $N$ represents the total number of frames.

### CTC Loss

The CTC loss is used to learn the alignment between the output sequence and the ground-truth text during the process of converting the generated speech into text via ASR. While the reconstruction loss ensures acoustic similarity by encouraging the generation of Mel-spectrograms that closely match the target, the CTC loss is designed to enforce alignment accuracy at the text level.

### Coefficient Tuning

In this study, the weights of the reconstruction loss and the CTC loss were predefined and kept fixed throughout training. The relative importance of the two loss terms was configured such that a higher weight was assigned to the reconstruction loss in order to prioritize speech reconstruction quality. This design choice reflects the consideration that participants may not be able to reproduce identical utterances consistently across both spoken speech and imagined speech conditions. Accordingly, stable Mel-spectrogram reconstruction was emphasized first and with text-level alignment via the CTC loss learned in a complementary manner.

## Generator Training

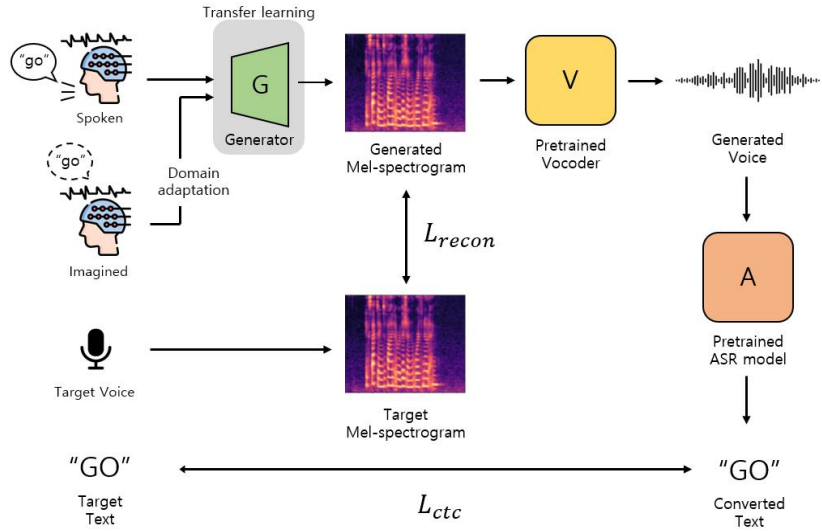

**Fig. 3| Training architecture of the EEG-to-Voice Generator.** Given preprocessed EEG signals as input, the Generator reconstructs Mel-spectrograms. The reconstructed Mel-spectrograms are compared with Mel-spectrograms converted from the target voice and are optimized using the acoustic reconstruction loss $L_{\text{rec}}$. In addition, the reconstructed Mel-spectrograms are synthesized into speech through a pretrained vocoder and then fed into a pretrained automatic speech recognition (ASR) model (A) to be converted into text. The decoded text is compared with the target text, and a CTC loss $L_{\text{ctc}}$ is applied for sequence-level training. The Generator is trained to simultaneously minimize both $L_{\text{rec}}$ and $L_{\text{ctc}}$.

Because the statistical properties and distributions of neural signals vary across subjects, this study trained an independent Generator for each participant. In addition, spoken speech and imagined speech exhibit inherently different neural signal characteristics. Spoken speech reflects overt articulation and includes muscle-related artifacts associated with speech production, whereas imagined speech involves no actual articulation and thus lacks muscle movements and direct speech-related acoustic manifestations. To account for these differences, separate Generators were trained for the two tasks.

Based on prior findings that neural representations related to spoken speech are partially reflected even during imagined speech, the Generator was first pretrained using spoken speech data. The pretrained weights were then used as initialization for imagined speech, and transfer learning–based domain adaptation was performed. This training strategy was motivated by the fact that, while the two tasks share the same targets and stimulus conditions, imagined speech data alone contain limited information due to the absence of actual speech signals. By leveraging representations learned from spoken speech as guidance signals, this approach compensates for the relatively weaker neural characteristics of imagined speech improves sample efficiency and enhances training stability.

Both neural signals and speech signals were segmented and used at the trial level. For neural signals, normalization was applied within each trial based on the minimum and maximum values to rescale the signals to a common range, and this preprocessing was performed to improve training stability and convergence behavior. The Generator was trained to take normalized neural signals as input and to output normalized Mel-spectrograms and an L1 loss was applied to minimize the difference between the generated Mel-spectrograms and those derived from speech recordings obtained during the spoken speech task.

In addition, the generated Mel-spectrograms were passed through a pretrained vocoder to synthesize speech which was then provided to an ASR module and a CTC loss was applied to encourage recognition of the target text. The final training objective of the Generator was defined as the simultaneous minimization of the L1 loss and the CTC loss, where the L1 loss served as a learning signal for acoustic reconstruction quality and the CTC loss enforced sequence-level alignment at the text level.

Each batch was composed exclusively of trials from a single subject to prevent inter-subject variability in neural signal distributions from being introduced during training. When acoustic similarity learning became unstable during training, the weighting coefficient of the L1 loss was reduced to relatively increase the contribution of text-level alignment thereby promoting training stability.

The Generator was trained for approximately 1,500 epochs with an initial learning rate of $1\times10^{-4}$. Optimization was performed using the AdamW optimizer with parameters $\beta_1 = 0.8$, $\beta_2 = 0.99$ and $\varepsilon = 1\times10^{-2}$. Training time varied depending on the amount of data, and using a single NVIDIA GeForce RTX 4090 GPU, training required approximately 48 hours per subject under the 2-s speech condition and approximately 84 hours per subject under the 4-s speech condition.

## Language Model setting

In this study, a minimal correction module based on Llama-3.2-1B-Instruct was designed to correct ASR outputs within a limited scope, including spelling errors, abnormal spacing, and simple token-level variations. The objective of this module is to preserve the semantic structure of the input sentence as much as possible, without introducing excessive additions or deletions of meaning. The design consists of four stages: (1) Input normalization, (2) Prompt-based candidate generation, (3) Candidate validity filtering and (4) Final selection based on a combined CER–PPL score. The language model used in this study was employed with it pretrained weights without any additional fine-tuning.

### Input Normalization

First, the ASR output text $x$ was converted into a normalized text $x_{\text{norm}}$ according to a set of predefined rules to ensure input consistency for the model.

$$x_{norm} = Normalize(x)$$

The normalization process consisted of converting all characters to uppercase, removing all characters other than alphabetic characters and whitespace, replacing commas and pipe symbols with spaces, and collapsing multiple consecutive spaces into a single space. This step was designed to reduce unnecessary token variations during the LLM-based candidate generation process and to establish a unified basis for comparing generated candidates.

Prompt-Based Candidate Generation
The normalized input text $x_{norm}$ was used to generate a set of five candidates $\{c_1, c_2, ..., c_5\}$ via beam search with a beam width of $B = 5$, based on a system prompt specifying the minimal correction rules. The LLM was instructed through the prompt not to add or delete words not to remove articles, and to restrict outputs to uppercase alphabetic characters and spaces only. The generated candidates were subsequently post-processed after prompt removal to convert them into a comparable text format for evaluation.

Candidate Validity Filtering
To enforce the principle of minimal correction, only candidates satisfying all the following conditions were retained in the final candidate set: (1) allowed character constraints, (2) restrictions on changes in word count, and (3) preservation of the set of articles. These filtering rules structurally prevent the LLM from attempting semantic reconstruction and play a critical role in maintaining alignment with the structure of the ASR output. In addition, the original text was always included in the candidate set to prevent cases in which the LLM-based correction process could degrade performance.

Candidate scoring
Candidate selection was performed using a combined score that integrates structural similarity to the input text, measured by CER, and linguistic naturalness, measured by perplexity (PPL).

1) CER
$$S_{CER}(c_k) = 1 - CER(c_k, x_{norm})$$
2) Perplexity
$$S_{PPL}(c_k) = 1/PPL(c_k)$$
3) Final
$$S(c_k) = 0.1 \cdot S_{CER}(c_k) + 0.9 \cdot S_{PPL}(c_k)$$

By assigning a higher weight to the perplexity term, the scoring function was designed to preserve semantic content while simultaneously promoting linguistic naturalness.

Final output selection
The final corrected output $y$ is defined as the candidate selected according to the scoring and filtering procedures described above.
$$y = argmax\ S(c_k)$$

The selected string is converted into a pipe-delimited token format and passed to the subsequent stages of the overall speech generation pipeline.

Language model–based correction is not included in the training of the EEG-to-Voice Generator. Unless otherwise specified, quantitative performance evaluations were conducted based on the results obtained prior to the application of the correction module.

## TTS Model Few-shot Fine-tuning

Speech generated by the Generator was fed into an automatic speech recognition (ASR) model and converted into text for subsequent language processing stages. A personalized text-to-speech (TTS) module was then applied as an auxiliary component for text-based speech resynthesis. Rather than training a TTS model from scratch, this study adopted a few-shot fine-tuning strategy using a limited amount of speech data.

For personalized speech synthesis, the xTTS-v2 model was employed, which supports multilingual speech synthesis and enables speaker voice cloning with a small amount of data. For subject-specific TTS fine-tuning, approximately 400 speech samples recorded during the experiments were used and these samples were split into training and test sets at a ratio of 9:1. During the fine-tuning process, a weighted-sum loss function was used considering the accuracy of the text-to–Mel-spectrogram mapping. The epoch with the lowest validation loss was selected as the optimal model, and early stopping was applied to prevent overfitting.

The personalized TTS module was not included in the training of the EEG-to-Voice Generator or in the primary quantitative performance evaluation. Unless otherwise specified, quantitative evaluations were conducted based on the results obtained prior to the application of the TTS module.

## Evaluation Metrics

In this study, both acoustic-level and linguistic-level metrics were employed to evaluate EEG-based speech reconstruction performance. All evaluations were conducted using one-to-one correspondence between the original speech and the reconstructed speech belonging to the same class and the same trial.

1) Mel-Cepstral Distortion (MCD)

    Mel-cepstral distortion (MCD) was used to quantify spectral distortion between the reconstructed speech and the reference speech[15]. First, WORLD-based spectral analysis was applied to both the original and reconstructed speech signals after which 24-dimensional mel-cepstral coefficients were extracted using the pysptk.sp2mc transformation. To restrict the evaluation to speech regions, energy-based voice activity detection (VAD) was applied to exclude silence frames. Because the two mel-cepstral sequences generally differ in length, temporal alignment was performed using dynamic time warping (DTW) based on Euclidean distance. The MCD value (in dB) was then computed from the distance between the aligned mel-cepstral vectors $c_t$ and $\hat{c}_t$ as follows.

    $$\text{MCD} = \frac{10}{\ln 10} \sqrt{2 \cdot \mathbb{E}[\|c_t - \hat{c}_t\|_2^2]}$$

2) Pearson Correlation Coefficient (PCC)

    The Pearson correlation coefficient (PCC) was used to evaluate how similar the reconstructed Mel-spectrogram was to the original Mel-spectrogram across the entire time–frequency domain. The two spectrograms were vectorized across both the temporal and Mel-frequency dimensions, and the Pearson correlation coefficient was then computed between the resulting vectors.

    n this study, the correlation was calculated directly using the full Mel-spectrogram values without applying any additional time-lag correction or separating voiced and unvoiced frames. Accordingly, PCC can be interpreted as a metric that reflects the overall similarity of global time–frequency patterns in the spectrograms.

3) Root Mean Square Error (RMSE)

    RMSE was used to evaluate the absolute magnitude difference between the reconstructed Mel-spectrogram and the original Mel-spectrogram. The squared error between the two spectrograms was computed across all time frames and Mel-frequency bins, and RMSE was calculated by taking the square root of the mean of these squared errors as follows:

$$\text{RMSE} = \sqrt{\frac{1}{TF}\sum_{t=1}^{T}\sum_{f=1}^{F}(x_{t,f}-\hat{x}_{t,f})^2}$$

4) Linguistic Accuracy (WER/CER)

Linguistic accuracy was evaluated based on text converted via ASR, using the word error rate (WER) and character error rate (CER). WER and CER assess how accurately the reconstructed speech preserves the original linguistic information independently of acoustic similarity metrics.

# Result

Trial-level accurate EEG-to-voice

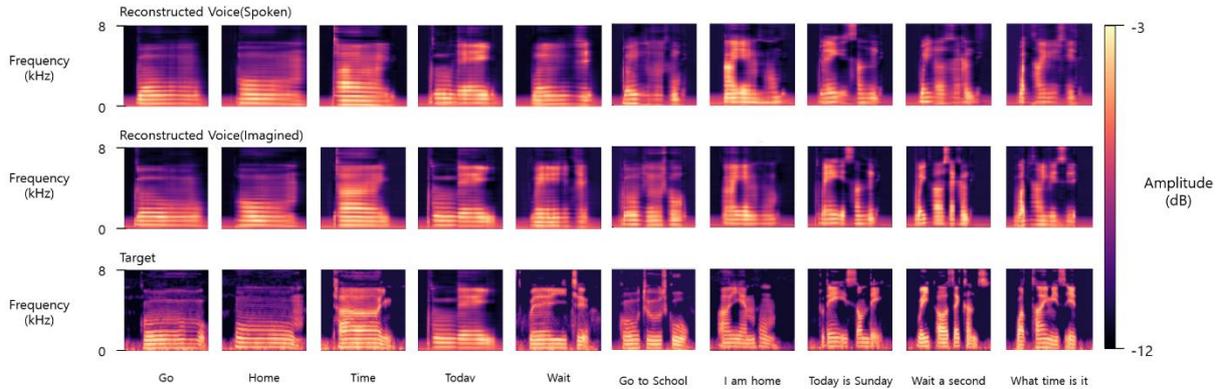

**Fig. 4| Mel-spectrogram examples under the 2-s speech condition.** Representative Mel-spectrograms obtained from open-loop reconstruction trials under the 2-s utterance condition are shown. The top row presents speech reconstructed from EEG signals during the spoken speech task, the middle row shows speech reconstructed from EEG signals during the imagined speech task, and the bottom row displays the corresponding target speech Mel-spectrograms. Each column corresponds to a different spoken word or sentence.

In the EEG-to-Voice paradigm, subject-specific Generators are used to directly reconstruct speech corresponding to the utterance length from trial-wise segmented EEG signals without DTW-based alignment. In this study, we signals during the spoken speech task, the middle row shows speech reconstructed from EEG signals during the imagined speech task, and the bottom row displays the corresponding target speech Mel-spectrograms. Each column corresponds to a different sentence, from left to right: "I am a student at this school," "I lost my notebook yesterday," "Please close the window before you go," "The library was very quiet," "We played baseball at the park," "He gave me information about the rule," "I saved document on the desktop," "I used a dictionary to find the word," "She waited for me near the airport," and "The classroom was empty when I arrived."

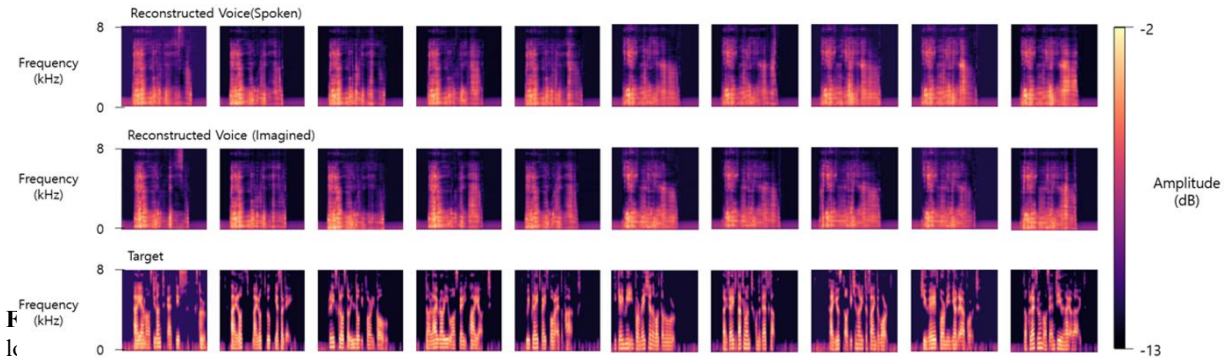

first evaluate how accurately the Generator reconstructs speech and subsequently analyze the linguistic accuracy when the reconstructed speech is converted into text via STT.

To this end, decoding performance was evaluated for both the spoken speech task and the imagined speech task under the 2-s speech condition using 10 classes that included both words and sentences, and the experiment was conducted with a total of 10 subjects (Table 1). In the proposed EEG-to-Voice paradigm, the Generator directly estimates the scale of the target Mel-spectrogram without normalization; therefore, evaluation was performed based on denormalized Mel-spectrograms. The Pearson correlation coefficient (PCC), which reflects time–frequency similarity between the reconstructed speech and the original speech, was $0.7406 \pm 0.0527$ for spoken speech and $0.7484 \pm 0.0522$ for imagined speech, indicating a high degree of similarity between the reconstructed and original Mel-spectrograms in both conditions (Fig. 4). The Root-mean-square error (RMSE), which measures absolute spectral differences, was $1.7136 \pm 0.2434$ for spoken speech and $1.7163 \pm 0.2308$ for imagined speech. In addition, the Mel-cepstral distortion (MCD), which quantifies spectral distortion, was $7.5530 \pm 0.3149$ for spoken speech and $7.5284 \pm 0.3200$ for imagined speech.

All acoustic similarity metrics, including PCC, RMSE, and MCD, exhibited consistent values with no substantial differences between spoken speech and imagined speech, suggesting that the proposed model achieves stable acoustic reconstruction performance regardless of speech production.

To evaluate linguistic accuracy, CER and WER were measured. For the spoken speech task, the CER was $0.2519 \pm 0.0539$ and the WER was $0.4466 \pm 0.1064$, while for the imagined speech task the CER was $0.2710 \pm 0.0459$ and the WER was $0.4748 \pm 0.0947$. Both CER and WER exhibited relatively similar levels across the two tasks indicating that the model preserves not only acoustic reconstruction performance but also text-based linguistic information to a certain extent.

|      | Spoken speech       | Imagined speech     |
| ---- | ------------------- | ------------------- |
| CER  | $0.2519 \pm 0.0539$ | $0.2710 \pm 0.0459$ |
| WER  | $0.4466 \pm 0.1064$ | $0.4748 \pm 0.0947$ |
| RMSE | $1.7136 \pm 0.2434$ | $1.7163 \pm 0.2308$ |
| MCD  | $7.5530 \pm 0.3149$ | $7.5284 \pm 0.3200$ |
| PCC  | $0.7406 \pm 0.0527$ | $0.7484 \pm 0.0522$ |

**Table. 1| Evaluation of decoding results under the 2-s speech condition.** Performance was evaluated for the spoken speech task and the imagined speech task using CER, WER, RMSE, MCD, and PCC, based on speech reconstructed through the EEG-to-Voice paradigm under the 2-s speech condition and the corresponding text decoded from the reconstructed speech. The results were calculated using a total of 100 samples per subject, including both word-level and sentence-level samples.

After examining the decoding results under the 2-s speech condition, decoding performance for the spoken speech task and the imagined speech task was further evaluated under an extended 4-s speech condition, and the experiment was conducted with a total of 12 subjects. As the utterance duration increased to 4 s, single words such as "baseball," "library," "notebook," "student," "window," "airport," "classroom," "dictionary," "document," and "information" were difficult to articulate continuously over the entire 4-s interval without intervening silent segments. For this reason, during both the training and inference stages of the model, these words were excluded from the original set of 20 classes, and evaluation was performed using only 10 classes composed of sentences.

|      | Spoken speech       | Imagined speech     |
| ---- | ------------------- | ------------------- |
| CER  | $0.2929 \pm 0.0668$ | $0.2474 \pm 0.0361$ |
| WER  | $0.4893 \pm 0.0948$ | $0.4346 \pm 0.0525$ |
| RMSE | $2.2323 \pm 0.2737$ | $2.2331 \pm 0.2605$ |
| MCD  | $9.4598 \pm 0.5542$ | $9.5125 \pm 0.6660$ |
| PCC  | $0.5841 \pm 0.0625$ | $0.5834 \pm 0.0584$ |

**Table. 2| Evaluation of decoding results under the 4-s speech condition.** Performance was evaluated for the spoken speech task and the imagined speech task using CER, WER, RMSE, MCD, and PCC, based on speech reconstructed through the EEG-to-Voice paradigm under the 4-s speech condition and the corresponding text decoded from the reconstructed speech. The

results were calculated using 100 sentence-level samples per subject.

Evaluation of acoustic reconstruction performance under the 4-s speech condition (Table 2) showed that the PCC was 0.5841 ± 0.0625 for the spoken speech task and 0.5834 ± 0.0584 for the imagined speech task, while the RMSE was measured as 2.2323 ± 0.2737 and 2.2331 ± 0.2605, respectively. In addition, the MCD was 9.4598 ± 0.5542 for spoken speech and 9.5125 ± 0.6660 for imagined speech.

PCC, RMSE, and MCD all exhibited consistent values with no substantial differences between the spoken speech and imagined speech tasks. However, compared to the 2-s speech condition, a decrease in acoustic similarity was observed for both spoken speech and imagined speech, with overall performance showing a downward trend (Fig. 5, Table 2).

To evaluate linguistic accuracy, CER and WER were measured. For the spoken speech task, the CER was 0.2929 ± 0.0668 and the WER was 0.4893 ± 0.0948, while for the imagined speech task, the CER was 0.2474 ± 0.0361 and the WER was 0.4346 ± 0.0525. Both CER and WER exhibited relatively similar levels across the two tasks, indicating that the reconstructed speech maintains a certain level of linguistic information stably even under the extended utterance-length condition.

Under the 4-s speech condition, the utterance length is extended to 4 s resulting in an increased temporal length of the Mel-spectrogram that must be reconstructed and consequently a higher learning difficulty. Taking this into account the loss function was designed to place greater emphasis on text decoding accuracy over the entire utterance rather than short-term Mel-spectrogram similarity. This training configuration accounts for the relatively lower acoustic similarity metrics observed and instead suggests that the model was trained with a stronger focus on reconstructing linguistic information.

Word position effects on character error rate

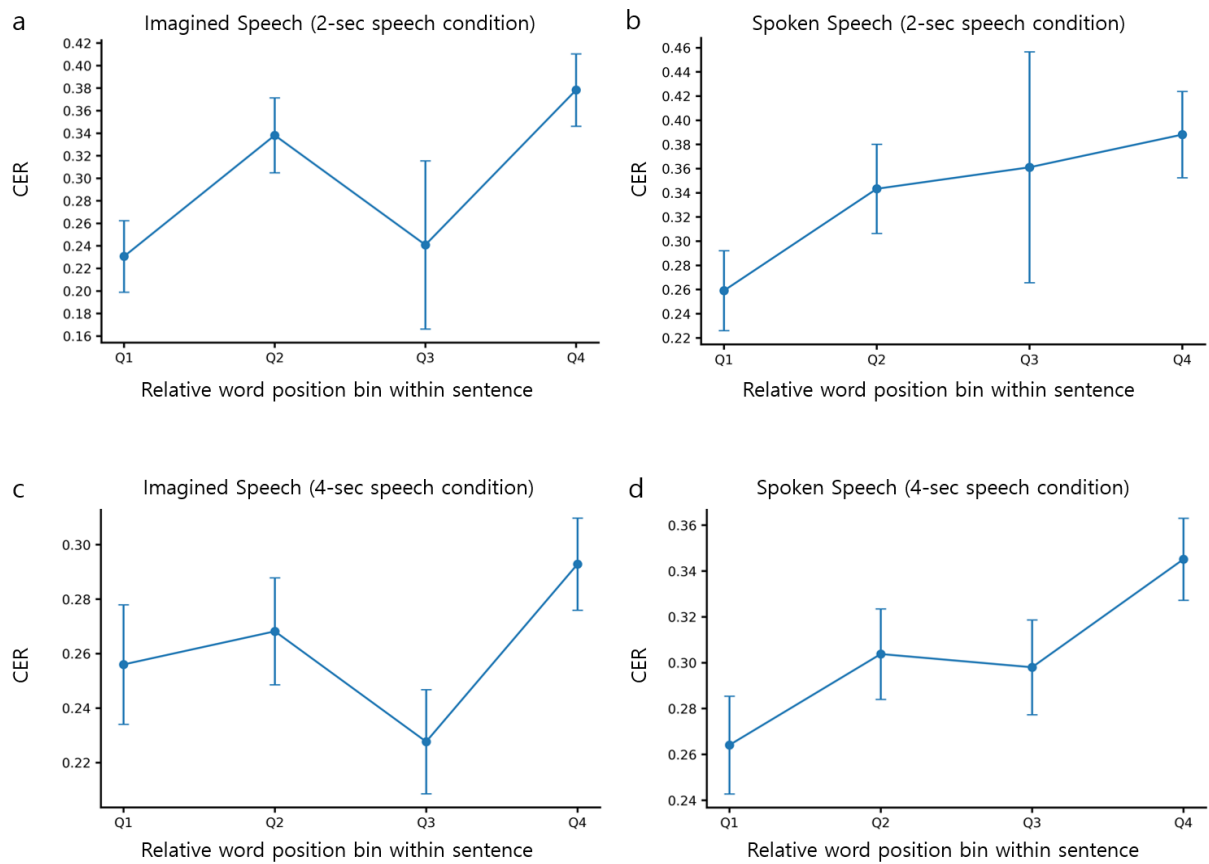

**Fig. 6| CER across relative word position bins under different speech conditions. a,** Imagined speech (2-s condition). **b,** spoken speech (2-s condition). **c,** imagined speech (4-s condition). **d,** spoken speech (4-s condition). Words were grouped into four equally spaced relative position bins within each sentence (Q1–Q4). Points denote sentence-level mean CER, and error bars indicate 95% confidence intervals across sentences.

|  | β | SE | P |
| --- | --- | --- | --- |
| Imagined (2 s) | 0.136 | 0.018 | <0.001 |
| Spoken (2 s) | 0.122 | 0.019 | <0.001 |
| Imagined (4 s) | 0.051 | 0.015 | <0.01 |
| Spoken (4 s) | 0.108 | 0.015 | <0.001 |

**Table. 3| Coefficients estimated using ordinary least squares with sentence-clustered robust standard errors.** Regression coefficients (β) were estimated using ordinary least squares with sentence-clustered robust standard errors. β denotes the change in CER per unit increase in relative word position within a sentence, and SE indicates the corresponding robust standard error. P values test the null hypothesis that the coefficient for word position equals zero. For the 2-s speech condition, regression analyses were performed using 50 sentence-level samples per subject (excluding word-only trials), whereas for the 4-s speech condition, 100 sentence-level samples per subject were used.

By training the EEG-to-Voice Generator under both the 2-s and 4-s speech conditions, decoded sentence-level text outputs were obtained for all subjects. Qualitative inspection of the decoding results revealed a tendency for character prediction accuracy to decrease toward the latter part of the sentence compared to the initial portion, with errors occurring more frequently in sentence-final words. To quantitatively evaluate these position-dependent error characteristics, changes in CER were analyzed at the word level as a function of the relative word position within each sentence.

Meanwhile, under the 4-s speech condition, all samples consisted of sentence-level samples allowing word-position interval analysis to be conducted without difficulty. In contrast, under the 2-s speech condition, word-level and sentence-level classes were mixed, making it difficult to define word-based position intervals evenly across four segments. Accordingly, to maintain analytical consistency, only samples corresponding to sentence-level classes were used for the analysis under the 2-s speech condition.

Fig. 6 illustrates the CER across relative word-position intervals within sentences for both imagined speech and spoken speech under the 2-s and 4-s speech conditions. Across all conditions, CER did not increase monotonically over the intervals; instead, localized variations were observed depending on speech type and utterance length.

Despite these non-monotonic patterns, word-level regression analysis accounting for sentence-level correlations (Table 3) revealed a weak but statistically significant positive association between word position and CER across all conditions ($\beta > 0$, $p < 0.01$). The magnitude of the position effect was larger under the 2-s speech condition than under the 4-s speech condition and among the 4-s conditions, it was weakest for imagined speech. These results suggest that, despite condition-specific localized fluctuations, decoding errors tend to increase toward the latter part of the sentence on average.

Meanwhile, the $R^2$ was generally low across conditions ($R^2 \leq 0.02$), indicating that word position alone explains only a limited portion of the overall variability in decoding errors.

Evaluation language model correction

By applying the EEG-to-Voice paradigm, performance was evaluated for directly reconstructing speech at the trial level from EEG signals without DTW-based alignment. Under both the 2-s and 4-s speech conditions, stable text decoding performance was confirmed for the spoken speech and imagined speech tasks.

However, because text decoding in this study was performed via ASR based on reconstructed speech, noise or acoustic distortions introduced during the speech reconstruction process may affect text decoding accuracy. To minimize these effects, final performance was evaluated after applying a language model–based text correction module that corrects errors within a limited scope such as spelling errors, abnormal spacing and simple token-level variations.

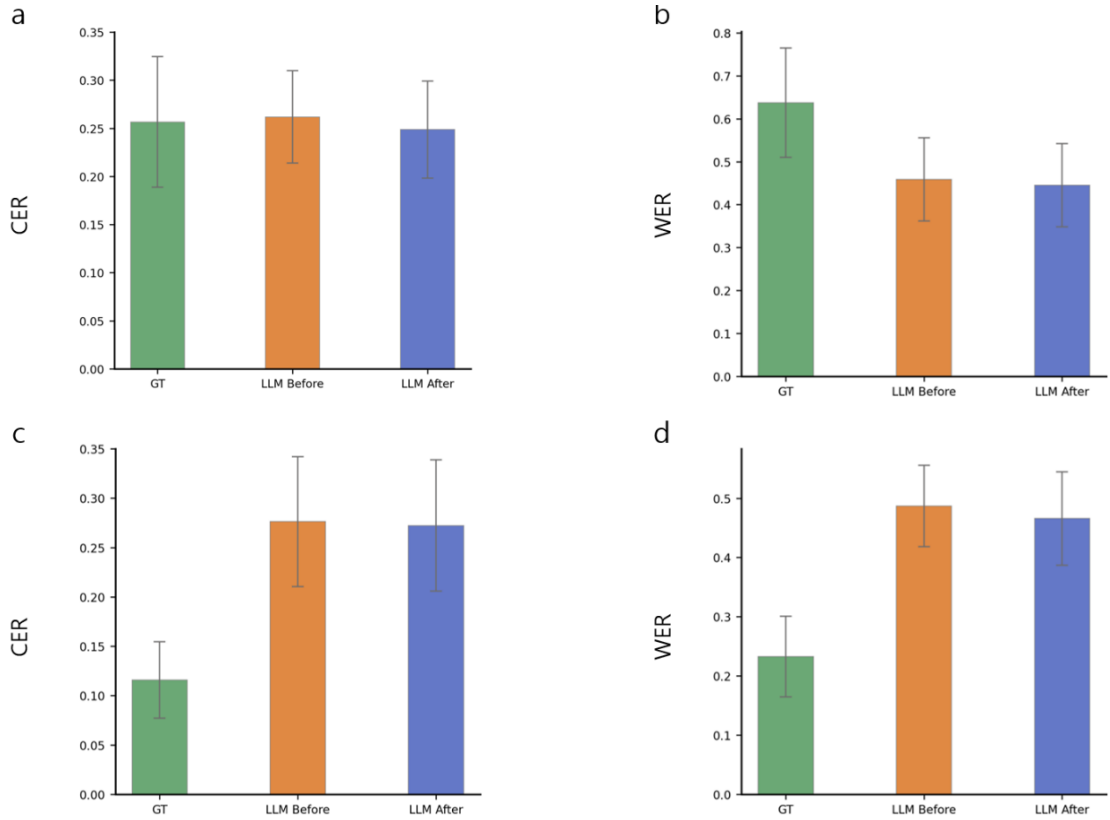

**Fig. 7| Text decoding accuracy before and after language model correction. a,** CER under the 2-s speech condition. **b,** WER under the 2-s speech condition. **c,** CER under the 4-s speech condition. **d,** WER under the 4-s speech condition. For all panels, Ground Truth (GT) denotes text decoding accuracy obtained by inputting spoken speech recordings directly into an ASR system. LLM Before indicates decoding accuracy obtained by applying ASR to speech reconstructed via the EEG-to-Voice paradigm, and LLM After indicates decoding accuracy after applying language model correction to the decoded text. Values for LLM Before and LLM After were computed by aggregating results from both imagined speech and spoken speech tasks. Bars represent mean values, and error bars denote standard deviation.

Fig. 7 compares the text decoding accuracy of outputs obtained through the EEG-to-Voice paradigm, including the Ground Truth (GT) with the text accuracy after applying language model–based correction. Figures 6a and 6b show the CER and WER under the 2-s speech condition, respectively while Figures 6c and 6d present the CER and WER under the 4-s speech condition.

|  | CER | WER |
| --- | --- | --- |
| Ground Truth (GT) | 0.2566 ± 0.0680 | 0.6375 ± 0.1275 |
| LLM Before | 0.2618 ± 0.0480 | 0.4586 ± 0.0968 |
| LLM After | 0.2487 ± 0.0504 | 0.4451 ± 0.0972 |

**Table. 4| Text decoding accuracy under the 2-s speech condition.** Text decoding accuracy was compared among the Ground Truth (GT), text decoded through the EEG-to-Voice paradigm (LLM Before), and text obtained after applying language model–based correction (LLM After) using CER and WER as evaluation metrics. Values are reported as mean ± standard deviation (mean ± s.d.). Under the 2-s speech condition, LLM Before was constructed based on decoding results from both the imagined speech task and the spoken speech task while LLM After reflects performance evaluated after applying language model–based correction to the same decoded outputs. The results were calculated using 100 samples per subject including both word-level and sentence-level samples.

|  | CER | WER |
| --- | --- | --- |
| Ground Truth (GT) | 0.1158 ± 0.0387 | 0.2327 ± 0.0679 |
| LLM Before | 0.2763 ± 0.0658 | 0.4870 ± 0.0699 |
| LLM After | 0.2723 ± 0.0665 | 0.4658 ± 0.0788 |

**Table. 5| Text decoding accuracy under the 4-s speech condition.** Text decoding accuracy was compared among the Ground Truth (GT), text decoded through the EEG-to-Voice paradigm (LLM Before), and text obtained after applying language model–based correction (LLM After) using CER and WER as evaluation metrics. Values are reported as mean ± standard deviation (mean ± s.d.). Under the 4-s speech condition, LLM Before was constructed based on decoding results from both the imagined speech task and the spoken speech task while LLM After reflects performance evaluated after applying language model–based correction to the same decoded outputs. The results were calculated using 100 sentence-level samples per subject.

Comparison of the Ground Truth (GT) results in Tables 4 and 5 shows that CER and WER under the 2-s speech condition are relatively higher than those under the 4-s speech condition. This can be attributed to differences in data composition between the two conditions. Under the 4-s speech condition, only sentence-level classes were used to eliminate silent segments allowing the utterance interval to be constructed without additional temporal extension. In contrast, under the 2-s speech condition, word-level and sentence-level classes were mixed, requiring words and sentences to be temporally extended to fill the utterance interval. As a result of these differences in data configuration, GT-based CER and WER were observed to be relatively higher under the 2-s speech condition compared to the 4-s speech condition.

Tables 4 and 5 present a comparison of text decoding accuracy between text decoded through the EEG-to-Voice paradigm (LLM Before) and text obtained after applying language model–based correction (LLM After). Under both speech conditions, CER and WER consistently decreased following the application of language model correction.

Under the 2-s speech condition, the CER decreased from 0.2618 ± 0.0480 to 0.2487 ± 0.0504, and the WER decreased from 0.4586 ± 0.0968 to 0.4451 ± 0.0972. Similarly, under the 4-s speech condition, the CER decreased from 0.2763 ± 0.0658 to 0.2723 ± 0.0665 and the WER decreased from 0.4870 ± 0.0699 to 0.4658 ± 0.0788. These results suggest that regardless of utterance length, language model–based correction within a limited scope can stably improve text decoding accuracy.